\documentclass[letterpaper, 10 pt, conference]{ieeeconf} 

\usepackage[utf8]{inputenc}
\IEEEoverridecommandlockouts 
\usepackage{cite}
\usepackage{blindtext}

\usepackage{booktabs}
\usepackage{arydshln} 

\usepackage{amsmath} 
\usepackage{bbm}
\usepackage{bm}
\usepackage{amssymb}
\usepackage{mathtools}

\usepackage{amsthm}

\usepackage{comment}

\theoremstyle{definition}

\newtheorem{assumption}{Assumption}

\newtheorem{problem}{Problem}

\usepackage{hyperref}
\hypersetup{
    colorlinks=true,
    linkcolor=black,
    filecolor=black,      
    urlcolor=blue,
    citecolor=blue,
    }
\usepackage{tikz}

\usepackage{todonotes}

\title{\LARGE \bf
LQ-OCP: Energy-Optimal Control for LQ Problems
}

\author{Logan E. Beaver, \emph{Member, IEEE} 
\thanks{L.E. Beaver is with the Department of Mechanical and Aerospace Engineering, Old Dominion University, Norfolk, VA USA (email: lbeaver@odu.edu).}%
}

\begin{document}
\maketitle
\thispagestyle{empty}

\begin{abstract}

This article present a method to automatically generate energy-optimal trajectories for systems with linear dynamics, linear constraints, and a quadratic cost functional (LQ systems).
First, using recent advancements in optimal control, we derive the optimal motion primitive generator for LQ systems--this yields linear differential equations that describe all dynamical motion primitives that the optimal system follows.
We also derive the optimality conditions where the system switches between motion primitives--a system of equations that are bilinear in the unknown junction time.
Finally, we demonstrate the performance of our approach on an energy-minimizing submersible robot with state and control constraints.
We compare our approach to an energy-optimizing Linear Quadratic Regulator (LQR), where we learn the optimal weights of the LQR cost function to minimize energy consumption while ensuring convergence and constraint satisfaction.
Our approach converges to the optimal solution $6,400\%$ faster than the LQR weight optimization, and that our solution is $350\%$ more energy efficient.
Finally, we disturb the initial state of the submersible to show that our approach still finds energy-efficient solutions faster than LQR when the unconstrained solution is infeasible.
\end{abstract}

\section{Introduction}

Motion planning is increasingly important with the growth and proliferation of robotics and cyber-physical systems in real-world environments \cite{Christensen2021AEdition}.
As these systems reach higher levels of autonomy, it is critical that they react quickly to their environment to guarantee safety and performance \cite{chalaki2021CSM}.
The need for performance must also be balanced with the energy cost of operating these systems.
Energy-aware control policies directly reduce the cost to power autonomous systems, and can often be implemented without sacrificing performance \cite{Notomista2023RelaxedBrushbots,Beaver2020AnFlockingb}.
Conserving energy also enables these systems to remain in the field longer \cite{Egerstedt2018RobotAutonomy}, and energy-efficient robots can be powered by smaller batteries, which reduces weight, increases efficiency, and decreases price.

Based on our recent results in optimal control \cite{Beaver2021DifferentialTime}, in this article we present a new approach to automatically generate optimal trajectories from a specified optimal control problem.
In particular, given a system with linear dynamics, linear constraints, and a quadratic cost functional (an LQ system), we derive a linear differential equation that is sufficient and necessary for optimality.
This ``motion primitive generator'' describes every dynamical motion that the system can achieve.
This is related to the dynamical motion primitives in the machine learning community \cite{dmp}, however, we can derive closed-form expressions for the dynamical motion primitives analytically because of the LQ structure.

One common motion planning technique is sample-based approaches, such as A*, RRT* or probabilistic road maps, to build a path that connects the initial and final state together \cite{campbell2020}.
These approaches tend to yield paths that are not energy efficient, as the path is constructed from a large number of line segments that meet at sharp angles \cite{beaver2023graph}.
Furthermore, the path is generally not constructed with a notion of energy-efficiency.
This can be somewhat mitigated by using a potential field, control barrier function, model predictive control, or a linear quadratic regulator to drive the system to a reference point in an energy-aware way \cite{campbell2020}.
These approaches can be used in conjunction with a path, i.e., having the agent track a moving reference, or they can be used to drive the agent to the desired state.
However, these approaches tend to focus on \emph{stability} rather than \emph{optimality}, and even if an energy-optimal tracking controller exists, it is only as efficient as the reference trajectory that it is tracking.

One of the most common approaches to optimize LQ problems is the Linear Quadratic Regulator (LQR).
LQR approaches drive the system to a reference state, and they have proven to be easy to implement, robust to noise and disturbances, and performant \cite{duriez2017machine}.
However, LQR suffers from two major drawbacks: 1) it cannot handle constraints on the state variables, and 2) the cost minimized by LQR controllers is a \emph{weighted tracking error} and has no notion of energy consumption or satisfying boundary conditions.
While it is is possible to construct an energy-minimizing LQR controller in some cases \cite{alavinasab2006active}, in general LQR methods cannot be applied to energy-optimal motion planning directly.
This is because every component of the state must appear in the LQR objective, whereas the actual energy consumption may be independent of some states, e.g., position.
Optimizing for the ``wrong'' objective function is a major pitfall of optimal control \cite{ledzewicz2022pitfalls}, as a solution can be optimal while still having poor performance.

In contrast, we present an approach to generate optimal trajectories for LQ systems that is energy-optimal, obeys state and control constraints, and is guaranteed to satisfy the prescribed (feasible) boundary conditions.
In fact, our approach solves a broader class of problems than LQR control--as we allow some states to have $0$ cost in the objective function.
Intuitively this makes sense; while LQR requires all states to be driven to the reference, we only seek to minimize the variables relevant to energy consumption in our proposed approach.

This work aligns with recent advances in optimal control algorithms.
One notable approach is the Method of Evolving Junctions (MEJ) \cite{li2017method,zhai2022method}, which similarly breaks the optimal control problem down into motion primitives.
However, where MEJ uses a global optimizer to determine the junctions between constraints, we solve the optimality conditions directly.
Furthermore, we provide the exact analytic form for all possible motion primitives, whereas MEJ solves for the motion primitives on a case-by-case basis.
Similarly, the NOnSmooth Numerical Optimal Control (NOSNOC) package \cite{NOSNOC} generates optimal solutions for non-smooth systems by exactly resolving the junctions between the different modes of the system.
However, where NOSNOC solves the optimal control problem via discretization, we resolve the dynamical motion primitives between junctions exactly.
In other words, rather than develop a general-purpose tool to solve a broad range of problems, we present an approach to automatically generate optimal trajectories from the specifications of an LQ problem.
We demonstrate this using Matlab's Symbolic Toolbox to construct our cost and constraint matrices, derive the optimal motion primitive generator, and solve for the particular optimal trajectory of our system on the order of seconds.
Best of all, the dynamical motion primitives can be computed and stored offline; this enables the optimal trajectory to be computed in real time--even for systems with large numbers of constraints \cite{beaver2023graph}.

The remainder of this article is organized as follows: We present the LQ optimal control problem in Section \ref{sec:problem}.
We derive the dynamical motion primitives in Section \ref{sec:dmp}, and discuss the junctions between primitives in Section \ref{sec:junctions}.
Numerical results for a submersible robot in a cave are presented in Section \ref{sec:simulation}; finally, we present conclusions and discuss some areas of future work in Section \ref{sec:conclusions}.

\section{Optimal Control Problem} \label{sec:problem}

We consider a linear dynamical system with state $\bm{x}\in\mathbb{R}^n$, control action $\bm{u}\in\mathbb{R}^m$, and dynamics,
\begin{equation}\label{eq:dynamics}
        \dot{\bm{x}} = A\bm{x} + B\bm{u},
\end{equation}
where $A, B$ are appropriately sized matrices.
The system is subject to linear constraints, which we denote
\begin{equation} \label{eq:constraints}
    C \bm{x} + D \bm{u} \leq \bm{e},
\end{equation}
where $C, D$ are $c\times n$ and $c\times m$ matrices, $\bm{e}$ is an $c\times 1$ vector, and $c$ is the number of constraints.
Our objective is to minimize a quadratic function of the state and control variables, i.e., 
\begin{equation} \label{eq:cost}
\begin{aligned}
    J(\bm{x}, \bm{u}) &= \bm{x}^{\intercal}Q\bm{x} + \bm{u}^{\intercal}R\bm{u} + 2\bm{x}^{\intercal}N\bm{u},
\\
    C(\bm{x}, \bm{u}) &= \int_0^T J(\bm{x}, \bm{u}) dt.
\end{aligned}
\end{equation}
Finally, we impose some working assumptions to solve this constrained linear-quadratic optimal control problem.

\begin{assumption} \label{smp:exist}
    The linear system \eqref{eq:dynamics} is controllable; the terminal time $T$ is large enough and the constraints \eqref{eq:constraints} admit the existence of a solution.
\end{assumption}

This article presents a numerical approach to generating optimal control trajectories in real time.
Thus, we derive our results under the premise that such a solution exists using Assumption \ref{smp:exist}.
One could relax this assumption by looking at the reachability of the boundary conditions as in \cite{Bryson1975AppliedControl}, however, this remains an open problem for systems with a large number of constraints.

\begin{assumption} \label{smp:symmetric}
    The matrix $Q$ in \eqref{eq:cost} is symmetric.
\end{assumption}
\begin{assumption} \label{smp:smooth}
    The matrix $R$ is full-rank.
\end{assumption}

Assumption \ref{smp:symmetric} is not fundamental for our analysis, it is only employed to simplify the partial derivatives that arise as part of the optimization.
Similarly, Assumption \ref{smp:smooth} ensures that the control trajectory is smooth and free of disconntinuities.
Furthermore, both assumptions are straightforward to implement in a real physical system.

To simplify our analysis, we assume that the system dynamics \eqref{eq:dynamics} are given in Brunovsk\'y normal form \cite{brunovsky}.
A transformation from a general linear system to Brunovsk\'y normal form always exists under Assumption \ref{smp:exist}.
This is a linear change in coordinates to an equivalent system of $m$ integrator chains of length $k_i$ with state $\bm{s}_i$ and action $a_i$, i.e.,
\begin{equation} \label{eq:brunovsky-dynamics}
    \dot{\bm{s}}_i = A_i\bm{s}_i + B_i a_i,
\end{equation}
with $A_i$ and $B_i$ matrices of the form,
\begin{equation}
    A = \begin{bmatrix}
        \bm{0}_{n_i} & I_{n_i\times n_i} \\
        0 & \bm{0}_{n_i}^{\intercal}
    \end{bmatrix}, \quad
    B = \begin{bmatrix}
        \bm{0}_{n_i} \\ 1
    \end{bmatrix},
\end{equation}
where $n_i = k_i-2$, $\bm{0}_{n_i}$ is a zero vector of length $n_i$, and $I_{n_i\times n_i}$ is the $n_i \times n_i$ identity matrix.

With the system in Brunovsk\'y normal form, we formulate a linear-quadratic optimal control problem (LQ-OCP).
\begin{problem} \label{prb:OCP}
Generate the trajectory $\bm{a}(t)$ that satisfies,
\begin{equation}
\begin{aligned} \label{eq:OCP}
    \min_{\bm{a}(t)} &~ \int_0^T \bm{s}^{\intercal}Q\bm{s} + \bm{a}^{\intercal}R\bm{a} + 2\bm{a}^{\intercal}N\bm{s} \\
    \text{subject to:} &~ \eqref{eq:brunovsky-dynamics}, \\
    & \bm{g}(\bm{s}, \bm{a}) = C\bm{s} + D\bm{a} - \bm{e} \leq 0, \\
    \text{given:} & ~ \bm{s}(0),\,  \bm{s}(T),
\end{aligned}
\end{equation}
where $T$ is a fixed time horizon that satisfies Assumption \ref{smp:exist} and $\bm{s}, \bm{a}$ are the stacked vectors of $\bm{s}_i$ and $a_i$, respectively.
\end{problem}
Note that the matrices $Q, R, N, C, D$ in Problem \ref{prb:OCP} are only equal to those in \eqref{eq:constraints} and \eqref{eq:cost} if the original system is in Brunovsk\'y normal form; otherwise, they are computed using the linear coordinate transformation.
Furthermore, while we assume the boundary conditions are all known in this work, it is relatively straightforward \cite{Beaver2021DifferentialTime} to generate additional boundary conditions if the initial or final value of a state is left free.

\section{Dynamical Motion Primitives} \label{sec:dmp}

To generate a solution to Problem \ref{prb:OCP} we apply Pontryagin's minimization principle following the standard procedure \cite{Bryson1975AppliedControl}.
We note that this is guaranteed to yield the optimal solution in this case, because Problem \ref{prb:OCP} is convex with linear constraints \cite{ledzewicz2022pitfalls}.
First, to simplify our notation, we define a single vector for each of the $i=1,2,\dots,m$ integrator chains,
\begin{equation} \label{eq:z}
    \bm{z}_i = \begin{bmatrix}
        \bm{s}_i \\ a_i
    \end{bmatrix}.
\end{equation}
For systems with integrator dynamics, i.e., those in Brunovsk\'y normal form, applying Pontryagin's minimization principle yields the ordinary differential equation \cite{Beaver2021DifferentialTime},
\begin{equation} \label{eq:ode}
    \sum_{n=0}^{k_i} (-1)^n \frac{d^n}{dt^n}\Big( \frac{\partial J}{\partial z_i^n} + \mu^{\intercal}\frac{\partial g}{\partial z_i^n} \Big) = 0,
\end{equation}
where $z_i^n$ corresponds to the $n$th row of the vector $\bm{z}_i$, i.e., the $n^{\text{th}}$ derivative of the base state $z_i^0$.
Note that $k_i$, i.e., the degree of each integrator chain, is the length of the vector $\bm{z}$.
We can write the optimality condition \eqref{eq:ode} as a product of vectors, namely, we define
\begin{equation} 
    \bm{D}_i \coloneqq \Bigg[1, -\frac{d}{dt}, \frac{d^2}{dt^2}, -\frac{d^3}{dt^3}, \dots, (1)^{k_i}\frac{d^{k_i}}{dt^{k_i}}\Bigg]^{\intercal},
\end{equation}
and let $\nabla_i$ denote the gradient operator with respect to $\bm{z}_i$.
Then, the optimality condition \eqref{eq:ode} is equivalent to,
\begin{equation} \label{eq:optimality-chain}
    \Big(\nabla_i J + \bm{\mu}^{\intercal}\nabla_i\bm{g} \Big) \bm{D}_i = 0.
\end{equation}
Note that the gradients of the cost and constraints are,
\begin{align}
    \frac{d}{d\bm{z}} J &= 2 \Big[ \bm{s}^{\intercal}Q + \big(N \bm{a}\big)^{\intercal},\, \bm{a}^{\intercal} R + \bm{s}^{\intercal}N \Big], \\
    \frac{d}{d\bm{z}} g &= \Big[C,\, D\Big],
\end{align}
To further simplify notation, we combine the constant matrices 
\begin{equation}
\begin{aligned} \label{eq:Kmatrix}
    K &= \begin{bmatrix}
        Q & N \\ N^{\intercal} & R
    \end{bmatrix}, \quad
    L &= \begin{bmatrix}
        C & D
    \end{bmatrix}.
\end{aligned}
\end{equation}
This implies the derivatives,
\begin{equation}
\begin{aligned} \label{eq:jg-derivatives}
    \frac{\partial}{\partial \bm{z}} J(\bm{z}) = 2\bm{z}^{\intercal} K, \quad
    \frac{\partial}{\partial \bm{z}} g(\bm{z}) = L.
\end{aligned}
\end{equation}
Finally, we can write \eqref{eq:optimality-chain} for each of the $i=1, 2, \dots, m$ integrator chains.
This yields $m$ ordinary differential equations,
\begin{align} \label{eq:optimality}
    \Bigg( 2\bm{z}^{\intercal}K_i + \bm{\mu}^{\intercal}L_i \Bigg)\bm{D}_i = 0,
\end{align}
where $K_i$, $L_i$ are the rows of \eqref{eq:jg-derivatives} corresponding to the states $\bm{s}_i$ and action $a_i$.
Subsequently, we analyze \eqref{eq:optimality} as a \emph{dynamical motion primitive generator} (e.g., as used in \cite{dmp}) to exhaustively calculate all possible motion primitives that the optimal trajectory consists of.

When no constraints are active the vector $\bm{\mu}(t) = \bm{0}$ by definition.
We re-interpret \eqref{eq:optimality} as the unconstrained dynamical motion primitive,
\begin{equation} \label{eq:unconstrained}
    \bm{z}^{\intercal}K_i \bm{D}_i = 0,
\end{equation}
Note that \eqref{eq:unconstrained} is a homogeneous linear ordinary differential equation of order $2 k_i$ with constant coefficients.
Differential equations with the form of \eqref{eq:unconstrained} are straightforward to solve and have a known exponential form.
The differential equation \eqref{eq:unconstrained} is of order $2|\bm{s}_i|$, thus the coefficients of the unconstrained solution can be solved using the $2|\bm{s}_i|$ boundary conditions.

In the constrained case, some subset of Lagrange multipliers are non-zero time-varying functions $\bm{\mu}(t)$.
Here, we only consider the elements $L$ that correspond to linearly independent active constraints.
Again, this results is a system of linear ordinary differential equations of order $2|\bm{s}_i|$; the Lagrange multipliers are a system of $c$ constraints can be solved to generate the function $\mu(t)$ in closed form. 
The result is a system of non-homogeneous linear differential equations with constant coefficients of order $2|\bm{s}|$,
\begin{equation}
\begin{aligned} \label{eq:constrained-mp}
    2\bm{z}^{\intercal}K_i\bm{D}_i &= -\mu^{\intercal}L_i\bm{D}_i, \\
    L\bm{z} - \bm{e} &\leq 0.
\end{aligned}
\end{equation}

\section{Constraint Junctions} \label{sec:junctions}

With the unconstrained and constrained dynamical motion primitives determined, the final challenge  is generating the optimal trajectory that satisfies the boundary conditions and remains feasible.
In this section, we assume a sequence of constraint activations is given, and demonstrate how the optimal trajectory is solved.
We address some approaches to select the sequence of constraints in following sub-sections.

The solution to Problem \ref{prb:OCP} is a piecewise collection of the unconstrained  \eqref{eq:unconstrained} and constrained \eqref{eq:constrained-mp} dynamical motion primitives.
We denote by $t_1$ the (unknown) time that the system switches between dynamical motion primitives, and refer to this time and the (unknown) state $\bm{s}(t_1)$ as a \emph{junction} between the two motion primitives.
At each junction, we have two possibilities:
\begin{enumerate}
    \item the constraint(s) becomes active only instantaneously at the junction, or
    \item the constraint(s) become active at the junction and remain active for some non-zero interval of time.
\end{enumerate}
In the following sub-sections we discuss how to automatically solve the optimality conditions at junctions in order to piece the optimal trajectory together.

\subsection{Tangency Conditions} 

It is critical to ensure that the system can transition through a junction without requiring an infinite impulse control input.
This is achieved through the so-called tangency conditions \cite{Bryson1975AppliedControl}, which ensure the system smoothly transitions between constraints by making the constraint an explicit function of the control input.
This is optimal under Assumption \ref{smp:smooth} \cite{Bryson1975AppliedControl}.

For a linear constraint,
\begin{equation}
    g = \bm{c}^{\intercal}\bm{x} + \bm{d}^{\intercal}\bm{u},
\end{equation}
where $\bm{c}^{\intercal}$ and $\bm{d}^{\intercal}$ are rows of the constraint matrices $C$ and $D$, we take derivatives until the coefficient of $\bm{u}$ is non-zero.
Taking a single derivative yields,
\begin{equation}
    \frac{d}{dt}g = \bm{c}^{\intercal}A\bm{x} + \bm{c}^{\intercal}B\bm{u}.
\end{equation}
We repeat this procedure until the product of $\bm{u}$ and its coefficient is non-zero, i.e.,
\begin{equation}
    \frac{d^q}{dt^q} = \bm{c}^{\intercal}A^{q}\bm{x} + \bm{c}^{\intercal}A^{(q-1)}B\bm{u}.
\end{equation}
Note that under Assumption \ref{smp:exist} the matrices $A$ and $B$ are controllable, thus the product $A^{(k)}B$ is guaranteed to have a non-zero product with $\bm{c}^{\intercal}$ for some $k\in\{0, 1, 2, \dots, n-1\}$.
Therefore, the general tangency condition and higher-order constraint with index $c$ is,
\begin{align}
    \bm{N}_c &= \begin{bmatrix}
        \bm{c}^{\intercal}\bm{s} - e_c, \\
        \bm{c}^{\intercal}A\bm{s}, \\
        \vdots\\
        \bm{c}^{\intercal}A^{(q-1)}\bm{s},
    \end{bmatrix}, \\
    g_c &= \bm{c}^{\intercal}A^{(q)}\bm{s} + \bm{c}^{\intercal}A^{(q-1)}B\bm{a}.
\end{align}
where $q$ is the minimum number such that $\bm{c}^{\intercal}A^{(q-1)}B$ is non-zero.

Finally, we note that if the constraint is only active \emph{instantaneously}, then the tangency conditions may not need to be satisfied--this is because the system immediately exits the constraint and so the constraint $g(\bm{s}, \bm{a})$ may have a corner.
In that case, the above analysis holds except the tangency matrix $\bm{N}_c$ only contains the first row.

\subsection{Optimality Conditions}

In the unconstrained case, the dynamical motion primitive \eqref{eq:unconstrained} presents a system of $2|\bm{s}|$ order differential equations that describe the optimal motion primitive.
There are $2|\bm{s}|$ boundary conditions provided in Problem \ref{prb:OCP}, which is sufficient to generate the unconstrained optimal trajectory.
If the unconstrained trajectory is infeasible, then the system must switch to a constrained motion primitive somewhere along the trajectory.
This can happen in two ways, 1) the system switches to a constrained motion primitive for a finite interval of time, or 2) the system switches to a constrained motion primitive only instantaneously.
Furthermore, the system may switch from one constrained motion primitive to another in the same way.

Let $t_1$ denote the (unknown) time of a junction where the system switches between two dynamical motion primitives.
To evaluate a variable $v$ at $t_1$ we write $v(t_1)$.
In general, the control actions and their derivatives may be discontinuous at $t_1$; using the standard notation we write $\bm{a}^-, \bm{a}^+$ to represent the left and right limits of a variable $\bm{a}$ evaluated at $t_1$.

In both cases the junction must satisfy a set of optimality conditions with respect to the dual vectors and the Hamiltonian \cite{Bryson1975AppliedControl}, i.e.,
\begin{align}
\bm{\lambda}^{\intercal^-} &= \bm{\lambda}^{\intercal^+} + \bm{\pi}^{\intercal}\frac{\partial \bm{N}}{\partial \bm{s}}, \label{eq:costateJump}\\
H^+ &= H^- + \bm{\pi}^{\intercal}\frac{\partial \bm{N}}{\partial t}, \label{eq:hamiltonianJump}\\
\frac{\partial H}{\partial \bm{a}^-} &= \frac{\partial H}{\partial \bm{a}^+} = \bm{0}. \label{eq:jumpOptimality}
\end{align}
The time-varying function $\bm{\lambda}$ is a costate vector, which is the dual variable to the state $\bm{s}$, and the function $H$ is the Hamiltonian, which is analogous to the Lagrangian for dynamical systems.
The variable $\bm{\pi}$ is a constant Lagrange multiplier; the elements of $\bm{\pi}$ are zero for rows of $\bm{N}$ that correspond to inactive constraints, otherwise they are (unknown) scalars that ensure constraint satisfaction at $t_1$.

For systems with integrator dynamics, the costates are equal to the partial sums of \eqref{eq:optimality}, i.e.,
\begin{equation} \label{eq:easy-lambda}
    \lambda_i^j = \sum_{n=1}^{k_i=j}(-1)^n\Bigg(\frac{\partial J}{\partial s_i^{j+n}} + \bm{\mu}^{\intercal}\frac{\partial g}{\partial y_i^{j+n}}\Bigg),
\end{equation}
where $\lambda_i^j$ is the covector for state $s_i^j$, i.e., the $j^{\text{th}}$ time derivative of base state $s_i^0$.
Writing \eqref{eq:easy-lambda} as a matrix equation yields,
\begin{equation} \label{eq:easy-lambda-lq}
    \lambda_i^j = \Bigg(\bm{z}^{\intercal}K_i + \mu^{\intercal}L_i\Bigg)\bm{D}_i^j,
\end{equation}
where $\bm{D}_{i}^j$ is $\bm{D}_i$ shifted down $j+1$ rows, and the first $j$ rows are equal to zero.
Furthermore, substituting \eqref{eq:easy-lambda-lq} into the jump conditions \eqref{eq:costateJump} yields an equation each base state $i = 1, 2, \dots, m$ and each derivative $j = 0, 2, \dots, |\bm{s}_i|$ ,
\begin{equation} \label{eq:jump}
    \Bigg(
    (\bm{z}^{\intercal^-} - \bm{z}^{\intercal^+}) K_i
    \Bigg)\bm{D}_i^j =
    \Bigg(
        (\bm{\mu}^{\intercal^-} - \bm{\mu}^{\intercal^+} ) L_i
    \Bigg)\bm{D}_i^j +
    \bm{\pi}\frac{\partial N}{\partial s_i^j}
\end{equation}
Note that, because the state $\bm{s}_i$ is continuous under Assumption \ref{smp:smooth}, for each integrator chain $i$ the condition \eqref{eq:jump} only determines the magnitude that the control action $a_i$ and its derivatives changes at the junction.
In particular, setting $j = k_i-1$ gives a condition on $a_i$, $j = k_i - 2$ gives a condition on $\dot{a}_i$ and $a_i$, etc.
This yields $|s|$ equations at the junction.

Next, combining \eqref{eq:hamiltonianJump} and \eqref{eq:jumpOptimality} yields a useful equation \cite{Beaver2021DifferentialTime},
\begin{equation} \label{eq:useful-jump}
    \Big(J^+ - J^-\Big) - \Big(\frac{\partial J}{\partial \bm{a}} + \mu^{\intercal} \frac{\partial \bm{g}}{\partial \bm{a}} \Big)^{\pm} \Big(\bm{a}^+ - \bm{a}^- \Big) = \bm{\pi} \frac{d\bm{N}}{dt}^{\mp}.
\end{equation}
Substituting in the definitions of $\bm{g}$ and $J$ using \eqref{eq:constraints} and \eqref{eq:cost} yields,
\begin{align}
    (\bm{a}^{\intercal}R\bm{a})^+ + (\bm{a}^{\intercal}R\bm{a})^- - 2\bm{a}^{\intercal^+}R\bm{a}^- \notag \\
    \pm \bm{\mu}^{\intercal^{\pm}}D\big(\bm{a}^+ - \bm{a}^-\big)
    = \bm{\pi} \frac{d \bm{N}}{d t}^{\mp}. \label{eq:hamjumplq}
\end{align}
This yields one more equation, which determines the unknown junction time $t_1$.

\subsection{Summary}

In the unconstrained case, the ordinary differential equation \eqref{eq:unconstrained} implies $2|\bm{s}_i|$ unknown constants of integration, which are solved exactly by the $2|\bm{s}_i|$ boundary conditions.
Because \eqref{eq:unconstrained} is a linear differential equation, the form of the unconstrained motion primitive can be solved analytically; then the particular trajectory coefficients can be solved online as a linear system of equations evaluated at $t=0$ and $t=T$.

If the unconstrained case is infeasible, then the system must encounter at least one \emph{junction}.
This allows the system to either switch between dynamical motion primitives or to activate a constraint instantaneously.
This introduces $2|\bm{s}| + c + 1$ unknowns; $2|\bm{s}|$ from the dynamical motion primitive \eqref{eq:constrained-mp}, $c$ for the time-varying Lagrange multipliers in the constrained motion primitive, and $1$ for the unknown time for the junction.
The corresponding equations are:
\begin{itemize}
    \item $|\bm{s}|$ equations from continuity in the state,
    \item $|\bm{s}|$ equations from the costate jump equations \eqref{eq:jump},
    \item $c$ equations from the active constraint $L_c\bm{z} = \bm{e}$, and
    \item $1$ equation from the Hamiltonan jump equation \eqref{eq:hamjumplq}.
\end{itemize}

We also note that if a constraint is active only instantaneously, then the tangency matrix has at most one row per constraint.
Furthermore, if the system is entering or exiting an unconstrained arc, then the corresponding Lagrange multipliers $\mu$ and the tangency vector $\bm{N}$ are zeros; this simplifies the optimality conditions significantly.

\subsection{Finding a Constraint Sequence}

From the previous sections we can generate the optimal trajectory given the sequence of constraint activations.
This is, in general, an open problem.
Many contemporary commercial solvers, e.g., NOSNOC \cite{NOSNOC} and MEJ \cite{li2017method}, do not explicitly address how the sequence of constraint activations is generated.
Some applications, e.g., minimum-time trajectories in polygonal environments \cite{zhai2022method}, admit a convex decomposition where the search space can be efficiently pruned using branch-and-bound.
We have found some success using a minimum distance heuristic to find the a feasible sequence of constraint activations \cite{beaver2023graph}, then branch-and-bound can be employed to prune the search space.

Other systems, such as connected and automated vehicles \cite{chalaki2021CSM}, have relatively low dimensionality and few constraints.
In this case, it generally efficient to try the unconstrained solution, check what constraint is violated, impose that constraint, and repeat until a solution is found.
However, this approach has neither proven to be efficient nor complete in general.
Furthermore, the approach scales poorly as the number of vehicles increases to an entire transportation network.

Along with the many pitfalls of applying optimal control in general \cite{ledzewicz2022pitfalls}, selecting the optimal sequence of constraint activations is crucial to ensure good performance.
This problem is largely unaddressed in the literature, and can reduce an approach from global to local optimality--particularly when many constraints are imposed on the system.
This is even true for the convex systems, including the LQ system analyzed in this work.

\section{Simulation Results} \label{sec:simulation}

\begin{figure}[ht]
    \centering
    \begin{tikzpicture}
    \node[circle,fill=black!70,minimum width=3pt,label={180:$(x,y)$}]
    (A) at (-2,0.75) {};
    \draw [-,ultra thick,color=red] (-3,0) -- (3,0);
    \draw [-,ultra thick,color=red] (-3,2) -- (3,2);
    \node () at (0,2.25) {$y=h$};
    \node () at (0,0.25) {$y=0$};

    \draw [->,ultra thick,color=black]
    (A) -- ++(1,0) node[right] {$u_x$};
    \draw [->,ultra thick,color=black]
    (A) -- ++(0,1) node[right] {$B$};
    \end{tikzpicture}
    \caption{A submersible robot at point $(x, y)$ with the (net) buoyancy force and forward thrust shown; horizontal lines correspond to the minimum and maximum height.}
    \label{fig:submersible}
\end{figure}
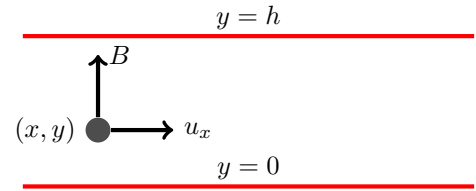

To study the performance of our control algorithm, we consider the problem of a simplified submersible robot navigating through a cave. 
This is depicted in Fig. \ref{fig:submersible}, where the submersible depicted with the relevant forces and height constraints.
The submersible has two control actions, applying forward thrust and changing its buoyancy.
The submersible's goal is to reach a goal state while minimizing its energy consumption and avoiding collisions with the floor and ceiling.

Applying Newton's second law yields a simplified submersible model,
\begin{equation}
    \begin{aligned}
        \ddot{x} &= -b_x \dot{x} + u_x, \\
        \ddot{y} &= -b_y \dot{y} + B,\\
    \end{aligned}
\end{equation}
where $b_x$ and $b_y$ are coefficients of drag, $u_x$ is the forward thrust, and $B$ is the buoyancy, which controlled via,
\begin{equation}
    \dot{B} = u_y.
\end{equation}
We seek to minimize energy consumption of the submersible robot with the objective function,
\begin{equation} \label{eq:ex-cost}
    J(\bm{x}, \bm{u}) = c_1 u_x^2 + c_2 u_y^2 + k_1 \dot{x}^2 + k_2 \dot{y}^2,
\end{equation}
where the first two terms minimize the actuation effort of thrusting and buoyancy change, and the last two terms penalize the power lost to drag.
We constrain the robotic submersible to remain within a minimum and maximum height and subject it to a minimum thrust constraint,
\begin{align}
    0 \leq y &\leq h \text{ (cavern height)}, \\
    u_x &\geq T_{\min} \text{ (minimum thrust)}.
\end{align}
To put the system in Brunovsk\'y normal form, we perform a  linear change of coordinates via feedback linearization,
\begin{align}
    u_x &= b_1 \dot{x} + a_x, \\
    \beta &= B - b_2\dot{y}, \\
    u_y &= a_y + b_2 \beta,
\end{align}
where $a_x, a_y$ are the new control actions and $\beta$ is the net upward acceleration.
The result is two integrator chains,
\begin{subequations}
\begin{align}
    \dot{x} &= v_x \quad \dot{v}_x = a_x, \\
    \dot{y} &= v_y \quad \dot{v}_y = \beta \quad \dot{\beta} = a_y.
\end{align}
\end{subequations}
Under the linear change in coordinates, the objective function \eqref{eq:ex-cost} becomes
$\bm{z}^{\intercal}K\bm{z}$, where, 
\begin{equation*}
    K = 
    \left[
    \begin{array}{ccccc:cc}
        0 & 0 & 0 & 0 & 0 & 0 & 0 \\
        0 & c_1 b_x^2+k_1 & 0 & 0 & 0 & 2 c_1 b_x & 0 \\
        0 & 0 & 0 & 0 & 0 & 0 & 0 \\
        0 & 0 & 0 & k_2 & 0 & 0 & 0 \\ 
        0 & 0 & 0 & 0 & c_2 b_y^2 & 0 & 2 c_2 b_y \\ \hdashline
        0 & 2 c_1 b_x & 0 & 0 & 0 & c_1 & 0 \\
        0 & 0 & 0 & 0 & 2 c_2 b_y & 0 & c_2
        \end{array}
    \right],
\end{equation*}
where
\begin{equation} \label{eq:ex-z}
    \bm{z} = \big[
    x, v_x, y, v_y, \beta, a_x, a_y
    \big].
\end{equation}
Furthermore, the constraints in the new coordinate frame are,
\begin{equation*} \label{eq:ex-constraint-matrix}
\bm{g} = 
    \left[
    \begin{array}{ccccc:cc}
        0 & 0 & -1 & 0 & 0 & 0 & 0 \\
        0 & 0 & 1 & 0 & 0 & 0 & 0 \\
        0 & -b_x & 0 & 0 & 0 & -1 & 0
        \end{array}
    \right]
    \bm{z}
    - 
    \begin{bmatrix}
        0 \\ h \\ -T_{\min}
    \end{bmatrix}
    \leq 0.
\end{equation*}
Note that this does not contain an explicit function of the control inputs in rows $1$ and $2$.
Thus, the corresponding tangency conditions $\bm{N}_1$ and $\bm{N}_2$ are,
\begin{equation*}
\begin{aligned}
    \bm{N}_1 &= 
    \begin{bmatrix}
        0 & 0 & -1 & 0 & 0 \\
        0 & 0 & 0 & -1 & 0 \\
        0 & 0 & 0 & 0 & -1
    \end{bmatrix} \bm{s},
    \\
    \bm{N}_2 &= 
    \begin{bmatrix}
        0 & 0 & 1 & 0 & 0 \\
        0 & 0 & 0 & 1 & 0 \\
        0 & 0 & 0 & 0 & 1
    \end{bmatrix} \bm{s} -
    \begin{bmatrix}
        h \\ 0 \\ 0
    \end{bmatrix},
    \end{aligned}
\end{equation*}

\subsection{Dynamical Motion Primitives}

Recall that we define the state $\bm{z}$ as \eqref{eq:ex-z}.
Thus, the sub-matrices $K_1$ and $L_1$ correspond to columns $1, 2,$ and $6$ while $K_2$ and $L_2$ correspond to columns $3, 4, 5,$ and $7$ of $K$ and $L$, respectively.
Substituting these values into the optimal motion primitive generator \eqref{eq:constrained-mp} for each integrator chain yields the dynamical motion primitives,
\begin{align}
    - 2(c_1 b_x^2 + k_1)a_x + 2 c_1 \ddot{a}_x &=  b_x \dot{\mu}_3 - \ddot{\mu}_3, \\
    - 2k_2\beta  + 2c_2 b_y^2\ddot{\beta} - 2c_2\ddddot{\beta}_y &= \dddot{\mu}_1 - \dddot{\mu}_2.
\end{align}

For the constrained case, note that rows $2$ and $3$ of the constraint matrix correspond to the minimum and maximum channel height, thus for $h > 0$ they cannot be active simultaneously or follow each other in a sequence.
This implies that out of the $2^c = 8$ possible dynamic motion primitives, only the following $6$ are physically feasible:

\begin{itemize}
\item Unconstrained Solution:
\begin{equation} \label{eq:ex-mp-1}
    \begin{aligned}
    a_x(t) &= \bm{c}_x e^{\kappa_1 t \pm \kappa_2}, \\
    \beta(t) &= \bm{c}_{y,1} e^{\kappa_3 t \pm \kappa_4}
     + \bm{c}_{y,2} e^{\kappa_5 t \pm \kappa_6}.
\end{aligned}
\end{equation}
\item Minimum Thrust:
\begin{equation} \label{eq:ex-mp-2}
    \begin{aligned}
    a_x(t) &= c_x e^{-bt}, \\
    \beta(t) &= \bm{c}_{y,1} e^{\kappa_3 t \pm \kappa_4}
     + \bm{c}_{y,2} e^{\kappa_5 t \pm \kappa_6}.
     \end{aligned}
\end{equation}
\item Height Constraint (2 cases):
\begin{equation} \label{eq:ex-mp-3}
    \begin{aligned}
    a_x(t) = \bm{c}_x e^{\kappa_1 t \pm \kappa_2}, \quad
    u_y = 0.
    \end{aligned}
\end{equation}
\item Fully Constrained (2 cases):
\begin{equation} \label{eq:ex-mp-4}
    \begin{aligned}
    a_x(t) = c_x e^{-bt}, \quad
    u_y = 0.
\end{aligned}
\end{equation}
\end{itemize}
Note that $\kappa_1, \kappa_2, \kappa_3, \kappa_4$ are known constants; $\bm{c}_x, \bm{c}_{y,i}$ are $2\times1$ vectors of unknown constants of integration, and $c_x$ is an unknown constant of integration.
We emphasize that while \eqref{eq:ex-mp-1}--\eqref{eq:ex-mp-4} are derived manually here, the dynamical motion primitives can be derived automatically using an automatic differentiation tool; we use the Matlab \texttt{Symbolic Math Toolbox} for this purpose.

For brevity we omit the calculation of the optimal jump conditions \eqref{eq:jumpOptimality} and we do not discuss how the sequence of constraint activations is selected.
The former can be computed automatically following the previous steps, while the latter is still an open problem in the literature.
Instead, we present a simulation results that compares the performance of our proposed approach to an energy-efficient LQR solution. 

\subsection{Unconstrained Solution}

The parameters used in our simulation study are given in Table \ref{tab:example-params}.
We selected these parameters to give a relatively high weight to actuation (through $c_1$ and $c_2$), while giving a higher penalty to energy lost to forward drag versus vertical drag (via $k_1$ and $k_2$). 
Note that when the submersible starts and ends at rest it stays within $p_y \in[p_y^0, p_y^f]$; similarly, for our travel time of $80$s, the travel distance of $100$ m is sufficiently large to keep the submersible's thrust above the minimum value.
Thus, the parameters in Table \ref{tab:example-params} represent a ``nominal'' scenario where the submersible follows an unconstrained trajectory.

\begin{table}[ht]
    \centering
    \begin{tabular}{ccccc}
         $c_1$,$c_2$ & $k_1$, $k_2$ & $b_x$,$b_y$ & $h$ & $T_{\min}$ \\
         $10$ & $(5,1)$ & $2.5$ & $25$ m & $1$ N/kg \\ \midrule
         $p_x^0$ & $v_x^0$ & $p_y^0$ & $v_y^0$ & $\beta^0$  \\
         $0$ m & $1$ m/s & $20$ m & $0$ m/s & $0$ N/kg \\ \midrule
         $p_x^f$ & $v_x^f$ & $p_y^f$ & $v_y^f$ & $\beta^f$  \\
         $100$ m & $1$ m/s & $1$ m & $0$ m/s & $0$ N/kg
    \end{tabular}
    \caption{Parameters, initial conditions, and final conditions used for the simulation study over a horizon of $T = 80$ seconds.}
    \label{tab:example-params}
\end{table}

\begin{figure}[ht]
    \centering
    \includegraphics[width=0.8\linewidth]{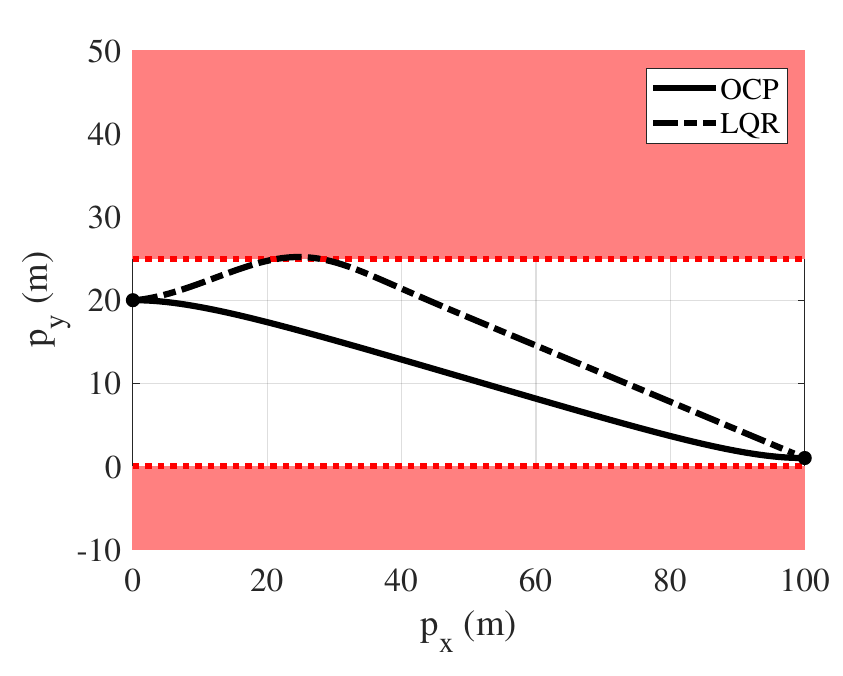}
    \caption{Resulting submerisble trajectories for the unconstrained case.}
    \label{fig:unconstrained}
\end{figure}

We present the unconstrained submersible trajectory in Fig. \ref{fig:unconstrained}.
Note that we do not use \eqref{eq:ex-mp-1}--\eqref{eq:ex-mp-4} in our simulation, rather we include the time for Matlab to generate and solve the differential equations using Matlab's \texttt{Symbolic Math Toolbox} on a desktop PC (i5 CPU @ 3.2Gz, 16 GB of RAM) using Matlab 2018b.
We start by constructing \eqref{eq:unconstrained} for each integrator chain; this takes approximately $1.3$ seconds.
Next, we solve the differential equations to generate the optimal motion primitives; this takes $0.375$ seconds for the general solution and $1.316$ seconds for the particular solution.
We take the difference to estimate the time required to apply the boundary conditions as $0.941$ seconds.
Finally, we convert the particular solution from a symbolic representation into a Matlab function using the \texttt{matlabFunction()} command--this takes approximately $11.3$ seconds, and it allows us to evaluate the state or control action of the vehicle at any instant of time along the trajectory.
Overall, we compute the optimal trajectory from Problem \ref{prb:OCP} in approximately 14 seconds--and most of the computations can be performed offline.

For the sake of comparison, we also generated a trajectory using an LQR controller \cite{duriez2017machine}.
LQR solves an unconstrained optimal control problem with an objective of the form \eqref{eq:cost}.
It seeks to do this by finding the optimal feedback matrix $K_{lqr}$ such that the optimal feedback law is,
\begin{equation}
    \bm{a} = -K_{lqr} \bm{s}.
\end{equation}
The feedback gain $K_{lqr}$ is time-varying for finite-horizon problems, but it can be approximated by the steady-state solution over large time horizons by solving the algebraic Ricatti equation \cite{duriez2017machine} that assume steady-state operation.
This can be achieved almost instantaneously using the \texttt{lqr()} command in Matlab, and is a standard approach in the literature.
However, for energy-aware robotic systems, there are two major caveats that come with LQR control:
\begin{enumerate}
\item LQR control cannot enforce \emph{constraints}--it only stabilizes the system about a reference.
\item LQR does not consider \emph{energy cost}--instead it minimizes a weighted sum of state errors.
\end{enumerate}

To find an energy-efficient LQR solution to the submersible problem, we constructed a new objective function $J_{lqr}$ with the form of \eqref{eq:cost}; this yields $28$ optimization variables that correspond to elements of the symmetric matrices $Q_{lqr}, R_{lqr}$ and the full matrix $N_{lqr}$.
To evaluate the performance, we simulated the LQR controller using the parameters in Table \ref{tab:example-params} and calculated the energy cost using \eqref{eq:cost}.
Finally, we penalized the LQR controller based on its distance from the desired final state, and we applied an infinite penalty if the matrices $Q_{lqr}, R_{lqr}, N_{lqr}$  were infeasible--i.e., did not admit a solution to the Riccati equation or violated the height constraints.
Note that we must solve this additional optimization problem because our cost matrix \eqref{eq:Kmatrix} does not admit an LQR solution!

The LQR optimization problem is non-linear, non-smooth, and high-dimensional; thus, we employed a genetic algorithm using Matlab's \texttt{Global Optimization Toolbox}.
We used the default parameters (200 population, 2,800 maximum iterations, 10 elites), and ran the algorithm until it converged after almost 1,000 iterations.
This required simulating the LQR system 196,480 times, and 68,958 of these evaluations were infeasible ($\approx 35\%$).
The total computational time was $914$ seconds, or approximately $15$ minutes--this is an order of magnitude larger than the $14$ seconds required for our proposed method on the same PC.

A comparison between our proposed approach and the LQR approach is presented in Fig. \ref{fig:unconstrained}.
While both approaches stay within the bounds, the LQR solution initially reaches the ceiling of the domain before diving to the final state.
We note that the LQR solution also does not reach the desired final state; this is because there is a  tradeoff between quickly converging to the final state versus minimizing energy consumption.
Again, this tradeoff explicitly appears in the LQR problem because it has no concept of energy-efficiency or constraints--only convergence.
The runtime and computational costs of this analysis are summarized in Table \ref{tab:unc_lqr}.

\begin{table}[ht]
    \centering
    \begin{tabular}{clll}
         & Proposed & LQR & Improvement \\ \toprule
         Computational Cost & $14.2$ seconds & $914$ seconds & $6,337\%$ \\
         Energy Cost        & $8,439$ units & $39,041$ units & $363\%$
    \end{tabular}
    \caption{Computational and energy cost for our proposed approach compared to energy-optimal LQR.}
    \label{tab:unc_lqr}
\end{table}

\subsection{Constrained Solution}

\begin{figure}[ht]
    \centering
    \includegraphics[width=.8\linewidth]{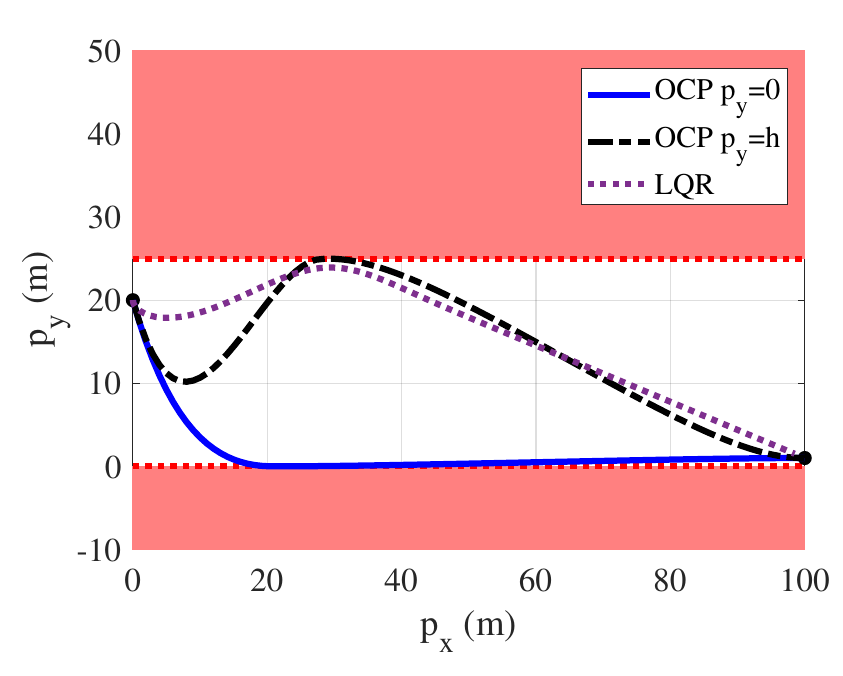}
    \caption{Resulting trajectories for the constrained case.}
    \label{fig:all-paths}
\end{figure}

\begin{figure}[ht]
    \centering
    \includegraphics[width=.8\linewidth]{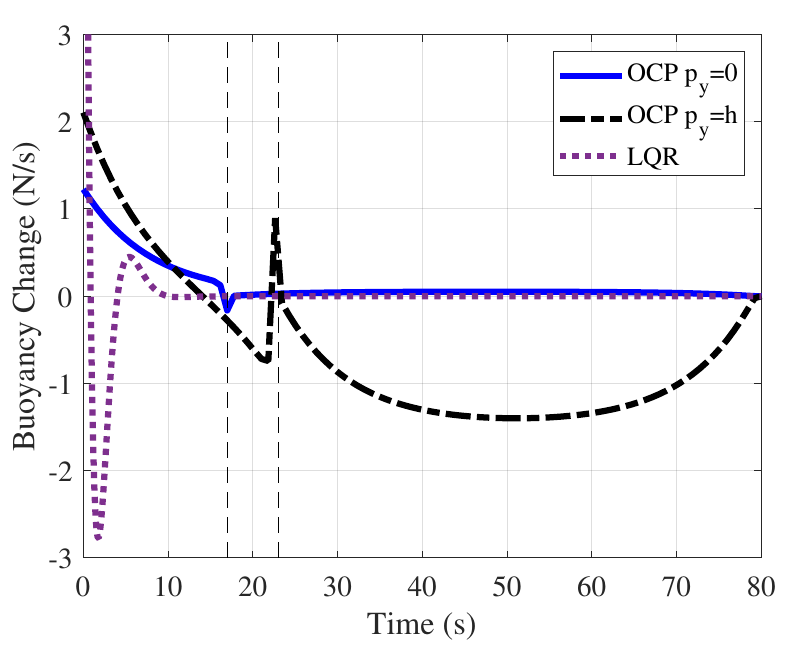}
    \caption{Change in buoyant force ($a_y$) for the optimal control and LQR solutions; the left vertical dashed lines demarcates the junction for the $p_y = 0$ case, and the right vertical dashed line demarcates the junction for $p_y = h$.}
    \label{fig:all-controls}
\end{figure}

The trajectory generated in the previous section remains within the domain $p_y\in[p_y^0, p_y^f]$ for the entire duration.
This is because the agent starts and ends at rest in the vertical direction, and overshooting the initial or final state would be energy-inefficient.
To force the system to overshoot the bounds of the domain, we adjust the initial velocity and net buoyancy of the submersible to 
    $v_y^0 = -3$ m/s, $\beta^0 = -1 N$.
This ensures that the unconstrained solution is infeasible, and requires us to follow the motion primitive $p_y = 0$ for some duration of the trajectory.
As the final height $p_y^f > 0$, the system will transition from an unconstrained to the height-constrained motion primitive at an unknown time $t_1$, then transition back to the unconstrained motion primitive at another unknown time $t_2 \geq t_1$.
Note that since the height constraint is independent of the integrator chain $[p_x, v_x, a_x]$, the control input $a_x$ and its derivatives are continuous via \eqref{eq:jump} and \eqref{eq:useful-jump}; thus the trajectory in the $x$ direction is the unconstrained solution.
The unconstrained and constrained solutions are depicted in Fig. \ref{fig:all-paths} and Fig. \ref{fig:all-controls}.

If $t_2 = t_1$, then the constraint is only active instantaneously and the system transitions between two unconstrained arcs.
In this case the tangency condition is just the constraint
$    \bm{N} = [ -p_y ]$.
Note that $v_y \leq 0$ just before the junction, as $p_y$ is decreasing; furthermore,  $v_y \geq 0$ just after the junction as $p_y \geq 0$.
Thus, it must hold that $v_y(t_1) = 0$.
Similarly, as $v_y$ achieves its minimum at $t_1$, its derivative $\dot{v}_y = \beta = 0$.
Next, note that \eqref{eq:jump} implies that either $a_y$ is continuous at $t_1$ or the jump in $a_y$ is equal to the jump in $\mu_3$.
Examining \eqref{eq:useful-jump} shows that the latter case leads to a contradiction, namely that the Lagrange multiplier $\pi = 0$.
Thus, $a_y$ is continuous at $t_1$ and $\mu(t_i) = \pi \neq 0$.
This yields the $7$ equations required to generate the optimal trajectory, i.e., continuity in the state, continuity of the control input, and $y(t_1) = v_y(t_1) = \beta(t_1) = 0$.
The resulting optimized trajectory is shown in Figs. \ref{fig:all-paths} and \ref{fig:all-controls}.

It is critical to realize that including the junction using the constraint $p_y = 0$ produces a \textit{candidate} optimal trajectory.
We selected this constraint using a common heuristic, i.e., we imposed the violated constraint as a motion primitive.
We could have used the motion primitive corresponding to $p_y = h$ at the junction; this solution is also shown in Fig. \ref{fig:all-paths} and \ref{fig:all-controls}.
This showcases a major open problem in constrained optimal control: 
selecting the wrong sequence of constraints can lead to feasible but sub-optimal solutions.
This is particularly true in cases where the cost of an action is linear, e.g., harvesting energy with regenerative braking.
Note that even the locally optimal $p_y = h$ solution is still a massive improvement over the LQR solution.

\begin{table}[ht]
    \centering
    \begin{tabular}{lcc}
         &  Energy Cost & Optimality Gap\\\toprule
     OCP with $v_y(t_1) = 0$ &  $8,458$ units & $0\%$ \\
     OCP with $v_y(t_1) = h$ &  $8,561$ units & $1.2\%$ \\
     LQR             & $41,832$ units & $395\%$ 
    \end{tabular}
    \caption{Energy cost for proposed and LQR solutions.}
    \label{tab:constrained}
\end{table}

\section{Conclusions and Outlook} \label{sec:conclusions}

In this article we derived an analytical solution that minimize the energy consumption of controllable LQ systems.
We derived the dynamical motion primitive generator \eqref{eq:constrained-mp}, and demonstrated how we can activate and deactivate individual constraints to generate all of the system's dynamical motion primitives offline.
We proved that the resulting motion primitives for an LQ system are linear ordinary differential equations--which always have a closed-form analytic solution.
Finally, we compared our proposed approach to an LQR solution and showed a massive improvement both in computational time and system performance.

The results of this article suggest a number of interesting avenues for future work.
First, finding optimal solutions to more general problems, e.g., using differential flatness \cite{Beaver2021DifferentialTime}, would broadly expand the applicability of these results.
Resolving existing computational bottlenecks, including an efficient methods to check for constraint violations and algorithms that yield a feasible sequence of motion primitives, would broadly expand the problems that are amenable to constrained optimal control.
Finally, converting multi-agent or game-theoretic problems into strategic-form games is an attractive research direction.

\bibliography{mendeley,refs,IDS_Pubs}

\end{document}